\documentclass[cameraready]{Interspeech}

\usepackage{cite}
\usepackage{amsmath,amssymb,amsfonts,bm}
\usepackage{graphicx}
\usepackage{textcomp}
\usepackage{xcolor}
\usepackage{multirow}
\usepackage{array}
\usepackage{listings}
\usepackage{subcaption}
\usepackage{xurl}

\def\Loss{\mathcal{L}}
\def\reward{r}
\def\IT{\mathrm{IT}}

\def\FDA{\mathrm{FDA}}

\def\old{\mathrm{old}}
\def\KLDiv{\mathcal{D}_{\mathrm{KL}}}
\def\clip{\mathrm{clip}}

\title{Aligning MusicLLM with Emotion \\using Instruction Tuning and Feedback-Driven Alignment}

\author[affiliation={1}, orcid=0009-0006-4843-9487]{Takuya}{Hasumi}
\author[affiliation={1}, orcid=0009-0007-5788-8193]{Welly}{Naptali}
\address{
    $^1$ LY Corporation
}

\email{}


\begin{document}

\maketitle

\begin{abstract}
This paper investigates whether music large language models (MusicLLMs) can be aligned for emotion regression.
While MusicLLMs have shown strong performance in music information retrieval tasks, their ability to predict arousal and valence scores remains limited, since emotion regression has not been an explicit training objective.
To examine whether MusicLLMs can be aligned with emotion, we train MusicLLMs on emotion regression and compare two strategies: instruction tuning and feedback-driven alignment.
Our experiments show that task-aware instruction tuning enables MusicLLMs to predict emotion levels to some extent, although the accuracy remains limited.
Applying feedback-driven alignment with a verifiable numerical reward substantially improves performance on both arousal and valence over instruction tuning alone.
We further show that our approach improves emotion regression performance while maintaining MusicQA capability.
\end{abstract}

\section{Introduction}
Multimodal large language models (multimodal LLMs) have rapidly advanced thanks to the recent growth of LLMs.
In the audio domain, integrating an audio encoder with an LLM has been explored for speech recognition, speech understanding, and general audio captioning~\cite{tang2023salmonn,ma2024embarrassingly,gong2024listen,chu2024qwen2,kong2024audio}.
Specifically, some MusicLLMs have been developed to address music information retrieval (MIR) tasks within a single model.
These models can describe the characteristics of a given music track and its musical instruments~\cite{doh2023lp,liu2024music,deng2023musilingo}.
A recent MusicLLM, Llark~\cite{gardner2023llark}, has even been trained to perform traditional MIR tasks such as key and tempo estimation.

% Issue
Given their success in various MIR tasks, MusicLLMs also appear well-suited to capture the intensity of emotion attributes (arousal/valence).
However, a recent benchmark~\cite{ma2025cmi} reveals that even state-of-the-art MusicLLMs are not good at emotion regression.
The performance is no better than a naive baseline that returns the dataset mean.
Nevertheless, despite this limitation in regression accuracy, MusicLLMs can jointly predict emotion levels and provide natural language explanations within a single model, which is difficult to achieve with traditional regression models alone.

% Training
We hypothesize that the low performance of MusicLLMs on the emotion regression task stems from a lack of explicit task-specific training.
Most MusicLLM benchmarks have excluded emotion regression from MIR task evaluation, except for \cite{ma2025cmi}.
Therefore, emotion regression training may have been overlooked and underexplored.
In addition, MusicLLMs have been trained to predict the next token, whereas the emotion regression task requires predicting continuous values.
Therefore, direct optimization of the score prediction error is necessary to produce reliable emotion level predictions.

% Solution
To investigate whether MusicLLMs can learn to capture emotional levels from audio through training, we train them on pairs of music tracks and corresponding emotion scores.
We adopt two training strategies: instruction tuning~\cite{wei2021finetuned}, which equips the model with basic regression capabilities using ground-truth scores, and feedback-driven alignment with verifiable numerical feedback~\cite{christiano2017deep}, which further refines the predictions.
Our experiments show that instruction tuning enables MusicLLMs to learn arousal estimation to some extent, but valence prediction remains partially challenging.
In contrast, feedback-driven alignment substantially improves both arousal and valence regression, demonstrating that feedback-driven alignment enables MusicLLMs to more accurately capture emotional perception in music.

Our contributions are summarized as follows:
\begin{itemize}
    \item
    We show that MusicLLMs can acquire emotion-regression capability through task-aware instruction tuning, although the accuracy is limited.
    \item
    We propose feedback-driven alignment using a verifiable numerical reward derived from regression error, and demonstrate that it substantially improves emotion regression performance, especially for valence.
    \item
    We demonstrate that our method improves regression performance while maintaining MusicQA capability, suggesting a step toward unified MIR systems.
\end{itemize}
\begin{figure}[tb]
    \centering
    \includegraphics[width=0.98\linewidth]{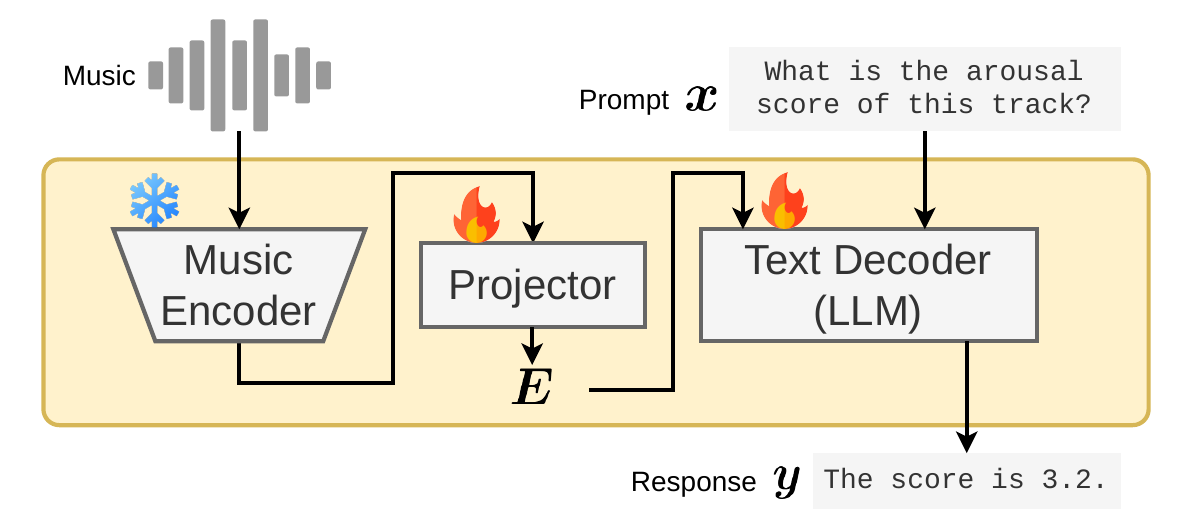}
        \caption{The overview of the music large language model (MusicLLM) for emotion regression.}
        \label{fig:proposed-method/framework}
\end{figure}

\section{Preliminary:\\Music Emotion Recognition}
Music emotion recognition (MER)~\cite{huron2000perceptual,yang2008regression} aims to estimate the subjective emotion of a music track.
This is essential to create mood-based playlist generation~\cite{bontempelli2022flow,kendre2023music} and emotion-based music generation~\cite{madhok2018sentimozart,neves2022generating,sulun2022symbolic}.

% Music emotion regression
Specifically, regression-based MER~\cite{yang2008regression} predicts arousal and valence scores, which correspond to two axes in Russell's circumplex model~\cite{russell1980circumplex}.
Arousal reflects the energetic level of the music track, while valence reflects its pleasantness.
Each score ranges from 1 to 9, where higher values indicate greater arousal or valence.
Unlike classification, the regression-based approach captures the continuous intensity of emotions.

% Conventional methods: encoder-based regression
To enable emotion regression, many encoder-based models have been proposed.
These models use convolutional neural networks (CNNs)~\cite{liu2017cnn}, recurrent neural networks (RNNs)~\cite{rajesh2020musical}, or Transformers~\cite{kang2025towards} to estimate emotions from audio features.
Some recent works~\cite{louro2024merge} even leverage lyric information by combining a text encoder with a music encoder.
However, no prior work has trained MusicLLMs to address this task, as mentioned earlier.

\section{Aligning MusicLLM with Emotion}
\subsection{Overview of MusicLLM}
\label{sec:methods/formulation}
\begin{figure*}[tb]
    \centering
    \includegraphics[width=0.65\linewidth]{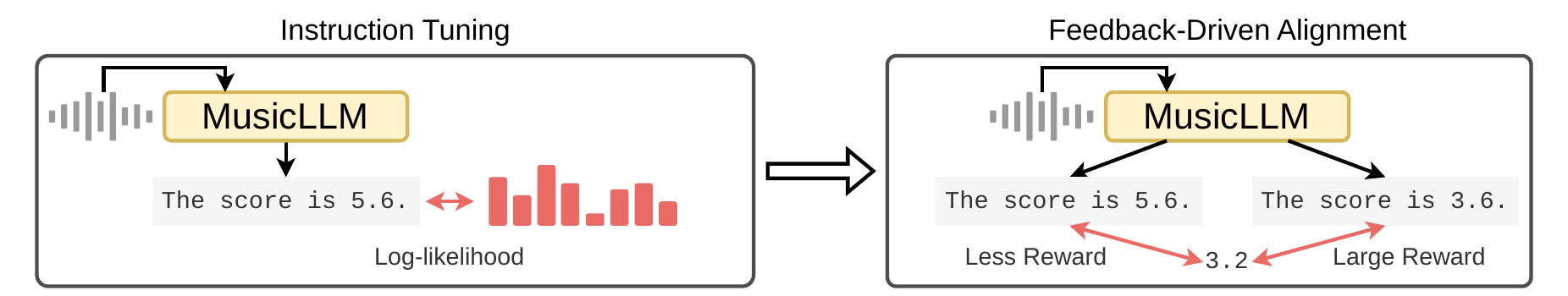}
    \caption{Training pipeline of MusicLLM in our work. First, the MusicLLM is trained with instruction tuning to learn the response format and a coarse emotion level. Then, the model is further fine-tuned by feedback-driven alignment to capture the fine-grained emotion level.}
    \label{fig:proposed-method/training-flow}
\end{figure*}
Our MusicLLM consists of three modules: a music encoder, a projector, and a text decoder (LLM), as illustrated in Fig.~\ref{fig:proposed-method/framework}.
Given a raw waveform, the music encoder produces a sequence of frame-level embeddings.
Then, the projector downsamples the sequence and maps it into the LLM token space, yielding latent embeddings $\bm{E}$ that are temporally aligned and dimension-matched to the text tokens.
The text decoder consumes both $\bm{E}$ and a textual prompt $\bm{x}$ and outputs a conditional distribution over response tokens $\bm{y}$.
The encoder is frozen during training, while the projector and decoder are optimized.

We adopt two training strategies: instruction tuning and feedback-driven alignment.
Fig.~\ref{fig:proposed-method/training-flow} outlines the overall training pipeline.
During instruction tuning, the model learns to map audio to arousal and valence scores from paired ground-truth annotations, establishing a basic regression capability.
Applying feedback-driven alignment after instruction tuning further refines the predictions using verifiable numerical feedback to capture fine-grained emotional levels.

\subsection{Instruction Tuning for Emotion Regression}
\begin{table*}[tb]
    \centering
    \caption{Examples of questions and answers for arousal score prediction.}
    \label{tab:methods/supervised-finetuning}
    \begin{tabular}{c|c}
        \hline
        Type & Sentence \\ \hline
        \multirow{2}{*}{Question} & On the 1-9 arousal scale, with low = relaxing and high = intense, what number fits this track? \\
        & How would you rate the music's valence—negative to positive—choosing a number between 1 and 9? \\ \hline
        Answer & The arousal level is \textit{\textless score\textgreater}. / I would assign a \textit{\textless score\textgreater} on the 1-9 scale. \\ \hline
    \end{tabular}
\end{table*}
Instruction tuning~\cite{wei2021finetuned} is a widely adopted approach for training MusicLLMs~\cite{liu2024music,gardner2023llark,deng2023musilingo}.
Instruction tuning teaches the model to predict the next token conditioned on the projected music embeddings $\bm{E}$ and a text prompt $\bm{x}$, with the training objective defined as the log-likelihood:
\begin{align}
    \Loss_{\IT}
    = \sum_{m=1}^{M}\log \pi_{\bm{\theta}}(y_{m}|\bm{E},\bm{x},\bm{y}_{<m}).
\end{align}
Here, $\bm{y}_{<m}$ denotes the subsequence of tokens preceding the $m$-th token.
We follow a chat-style format:

\begin{lstlisting}[label={lst:chat-template}, basicstyle=\small\ttfamily,frame=single,aboveskip=2pt,belowskip=2pt]
 USER: <question>
 ASSISTANT: <response>
\end{lstlisting}
In this format, \texttt{\textless response\textgreater} corresponds to $\bm{y}$ and the preceding part corresponds to $\bm{x}$.

Since existing emotion regression datasets~\cite{aljanaki2017developing,louro2024merge} contain only numeric scores, we generated pseudo-QA pairs for score prediction.  
Table~\ref{tab:methods/supervised-finetuning} shows some examples.  
Using GPT-4o~\cite{hurst2024gpt}, we designed 20 question templates and 20 answer templates to ensure consistent formatting and enable natural-language outputs of continuous ratings.
The scores are programmatically filled with one decimal place, such as ``5.6''.

\subsection{Feedback-Driven Alignment for Emotion Regression}
\label{sec:methods/feedback-alignment}
Feedback-driven alignment~\cite{christiano2017deep} refines the output of LLMs to maximize task-aware rewards. In this work, we use the term feedback-driven alignment (FDA) to refer to policy optimization using verifiable numerical rewards.
This method has been shown to improve mathematical reasoning in LLMs~\cite{shao2024deepseekmath} and the performance of visual question answering in vision-language models~\cite{sun2023aligning}.

Several methods have been proposed for feedback-driven alignment, including proximal policy optimization (PPO)~\cite{schulman2017proximal} and group relative policy optimization (GRPO)~\cite{shao2024deepseekmath}.
While PPO can be applied to this task, it requires learning a value function (critic), which can increase training complexity and often results in unstable optimization.
Therefore, we adopt GRPO for emotion regression with MusicLLMs, since it optimizes the policy using scalar regression rewards and avoids value-function learning, leading to a simpler and typically more stable training\footnote{Preference optimization methods such as direct preference optimization~\cite{rafailov2023direct} can be applied by constructing pairwise preference comparisons from scalar rewards. However, since our regression setting provides a scalar reward, we directly utilize policy optimization.}.

In GRPO for MusicLLMs, we first generate $G$ responses for a pair of $\bm{E}$ and $\bm{x}$:
\begin{align}
    \hat{y}_{g,m}\sim\pi_{\bm{\theta}}(\cdot|\bm{E},\bm{x},\hat{\bm{y}}_{g,<m})~~~\mathrm{for}~m=1,\ldots,M_{g},
\end{align}
where $g$ is the index of the $G$ generated samples, and $M_{g}$ is the number of tokens in the $g$th sample.
Using the set of generated responses $\{\hat{\bm{y}}_{g}\}$ and the ground-truth response $\bm{y}$, the advantage $A_{g}$ is computed as
\begin{align}
    A_{g} = \frac{r(\bm{x},\bm{y},\hat{\bm{y}}_{g}) - \mu(\bm{x},\bm{y},\hat{\bm{y}}_{1},\ldots,\hat{\bm{y}}_{G})}{\sigma(\bm{x},\bm{y},\hat{\bm{y}}_{1},\ldots,\hat{\bm{y}}_{G})},
\end{align}
where $r(\bm{x},\bm{y},\hat{\bm{y}}_{g})$ denotes a user-defined reward function.
$\mu(\bm{x},\bm{y},\hat{\bm{y}}_{1},\ldots,\hat{\bm{y}}_{G})$ and $\sigma(\bm{x},\bm{y},\hat{\bm{y}}_{1},\ldots,\hat{\bm{y}}_{G})$ denote the mean and standard deviation of $r(\bm{x},\bm{y},\hat{\bm{y}}_{g})$, respectively.

The objective function is defined as
\begin{align}
    \Loss_{\FDA}
    &= \frac{1}{\sum_{g=1}^{G}M_{g}}\sum_{g=1}^{G}\sum_{m=1}^{M_{g}}(\bar{A}_{g,m} - \beta\Delta_{g,m}), \\
    \bar{A}_{g,m}
    &= \min\left(\lambda_{g,m}A_{g},\clip(\lambda_{g,m}, 1 - \epsilon, 1 + \epsilon)A_{g}\right), \\
    \lambda_{g,m}
    &= \frac{\pi_{\bm{\theta}}\left(\hat{y}_{g,m}\middle|\bm{E},\bm{x},\hat{\bm{y}}_{g,<m}\right)}{\pi_{\old}\left(\hat{y}_{g,m}\middle|\bm{E},\bm{x},\hat{\bm{y}}_{g,<m}\right)}, \\
    \Delta_{g,m}
    &= \KLDiv\left[\pi_{\bm{\theta}}(\cdot|\bm{E},\bm{x},\hat{\bm{y}}_{g,<m})\middle||\pi_{\mathrm{ref}}(\cdot|\bm{E},\bm{x},\hat{\bm{y}}_{g,<m})\right],
\end{align}
where $\KLDiv\left[\cdot\middle||\cdot\right]$ is the Kullback--Leibler divergence between probability distributions, and $\beta$ is the KL penalty to reduce the discrepancy from the reference distribution $\pi_{\mathrm{ref}}$ (i.e., the model after instruction tuning).
$\epsilon > 0$ is a tiny value that controls the clipping range.
$\clip(\cdot,v_{L},v_{H})$ clips values to the range $[v_{L}, v_{H}]$.

For the reward function, we define:
\begin{align}
    \reward(\bm{x}, \bm{y}, \hat{\bm{y}})
    &= \begin{cases}
        - 200 & \mathrm{(if~score~parsing~error)} \\
        - (s(\hat{\bm{y}}) - s(\bm{y}))^{2} & \mathrm{(otherwise)}
    \end{cases},
    \label{eq:methods/dpo/reward}
\end{align}
where $s(\cdot)$ extracts the arousal or valence score from the generated text\footnote{For example, if ``\textit{I would assign a 3.6 on the 1-9 scale.}'' is given as a text, the score is extracted as the numerical value of 3.6.
We implemented $s(\cdot)$ to avoid extracting unrelated numbers such as 1 or 9 in this example.}.
Our reward function imposes a large penalty ($-200$) for parsing failures to encourage consistent formatting, and otherwise yields the negative squared error as the reward.

\section{Experiments}
\begin{table*}[tb]
    \centering
    \caption{Performance comparison for music emotion regression under different training strategies, excluding general music QA. IT and FDA denote instruction tuning and feedback-driven alignment, respectively.}
    \label{tab:experiments/results-without-musicqa}
    \begin{tabular}{c|c|c|cc}
        \hline
        \multirow{2}{*}{Model} & IT & FDA & \multicolumn{2}{c}{$R^{2}$ (arousal/valence)} \\
        & DEAM \& MERGE & DEAM \& MERGE & DEAM & MERGE \\ \hline
        \multirow{2}{*}{MusicFM + Vicuna} & $\checkmark$ & & $0.38$/$0.26$ & $0.40$/$0.05$ \\
        & $\checkmark$ & $\checkmark$ & $0.56$/$0.55$ & $\bm{0.55}$/$0.29$ \\ \hline
        MusicFM probing & --- & --- & $\bm{0.62}$/$0.31$ & $0.51$/$\bm{0.43}$ \\ \hline
        Encoder-based models & --- & --- & $0.52$/$\bm{0.62}$~\cite{kang2025towards} & $0.48$/$0.31$~\cite{louro2024merge} \\ \hline
    \end{tabular}
\end{table*}
\begin{table*}[tb]
    \centering
    \caption{Performance comparison for music emotion regression under different training strategies. B-U, M-R, and R-L denote the BLEU@4, METEOR, and ROUGE$_{\mathrm{L}}$ scores. MusicQA contains general music QA without emotion regression.}
    \label{tab:experiments/results-with-musicqa}
    \setlength{\tabcolsep}{4pt}
    \begin{tabular}{c|cc|c|cc|ccc}
        \hline
        \multirow{2}{*}{Model} & \multicolumn{2}{c|}{IT} & FDA & \multicolumn{2}{c|}{$R^{2}$ (arousal/valence)} & B-U & M-R & R-L \\
        & MusicQA & DEAM \& MERGE & DEAM \& MERGE & DEAM & MERGE & \multicolumn{3}{c}{MusicQA} \\ \hline
        \multicolumn{1}{l|}{\textit{Ours}} & & & & & & \\
        \multirow{3}{*}{\shortstack{MusicFM \\ + Vicuna}} & $\checkmark$ & & & $-0.16$/$-0.33$ & $-0.29$/$-0.19$ & $0.13$ & $0.14$ & $0.38$ \\
        & $\checkmark$ & $\checkmark$ & & $0.32$/$-0.35$ & $0.43$/$0.01$ & $0.15$ & $0.15$ & $0.40$ \\
        & $\checkmark$ & $\checkmark$ & $\checkmark$ & $\bm{0.48}$/$\bm{0.35}$ & $\bm{0.50}$/$\bm{0.24}$ & $0.15$ & $0.15$ & $0.39$ \\ \hline
        \multicolumn{1}{l|}{\textit{Open models}} & & & & & & & & \\
        Qwen2-Audio & --- & --- & --- & $-3.47$/$-2.02$ & $-2.63$/$-0.48$ & $0.07$ & $0.12$ & $0.27$ \\
        Phi-4-Multimodal & --- & --- & --- & $-2.22$/$-3.52$ & $-2.42$/$-0.74$ & $0.10$ & $0.13$ & $0.38$ \\ \hline
    \end{tabular}
\end{table*}

\subsection{Datasets}
To evaluate whether MusicLLMs can learn emotion levels from music, we used two publicly available emotion regression datasets: DEAM~\cite{aljanaki2017developing} and MERGE~\cite{louro2024merge}. 
DEAM consists of about 45-second clips with arousal and valence scores.
We follow the training/validation/test split used in \cite{kang2025towards}, resulting in 1261/271/270 samples.
MERGE consists of about 30-second clips labeled semi-automatically by mapping AllMusic\footnote{\scriptsize\url{https://www.allmusic.com/}} tags through Warriner’s dictionary~\cite{warriner2013norms}, yielding 2490/532/532 samples.
We also evaluate on MusicQA~\cite{liu2024music}, a dataset of open-ended questions about music, instruments, and mood, to assess general music QA ability.
For all experiments, audio was partitioned into 30-second segments: random segments were cropped during training, and the middle segment was used for evaluation.

\subsection{Proposed method}
Our MusicLLM is based on the SLAM-LLM~\cite{ma2024embarrassingly} architecture, consisting of a music encoder, a text decoder, and a projector as described in Sec.~\ref{sec:methods/formulation}.

\noindent\textbf{Music encoder}: We used MusicFM~\cite{won2024foundation}, pretrained on the Million Song Dataset (MSD)~\cite{bertin2011million}.
\noindent\textbf{Text decoder (LLM)}: We adopted Vicuna~\cite{wei2023vicuna}, a 7B-parameter open-source LLM trained on large-scale text corpora.

\noindent\textbf{Projector}: It consists of two linear layers with a ReLU in between.
Inputs are concatenated along the temporal axis at a downsampling rate of $5$, projected to an intermediate space, and mapped to the decoder dimension.

During training, we froze the parameters of the music encoder.
We applied low-rank adaptation (LoRA)~\cite{hu2022lora} to the query and value projection matrices in the text decoder, with rank $8$.
Each model was trained on a single NVIDIA A100 GPU, with a maximum response length of $128$ tokens.

In instruction tuning, the model is trained for $50$ epochs with the AdamW~\cite{loshchilov2017decoupled} optimizer, a batch size of $8$, and gradient accumulation over $2$ steps.

In feedback-driven alignment, the model is trained for $10$ epochs with a batch size of $1$ using gradient accumulation over $2$ steps.
GRPO parameters were set to $G=8$, $\beta=0$, and $\epsilon=0.02$.
We chose $\beta=0$ since recent LLM studies~\cite{hu2025open,yu2025dapo} show that removing the KL divergence term stabilizes training and improves performance.

\subsection{Baselines}

\noindent\textbf{MusicFM probing}: We applied layer normalization to the output of the music encoder (MusicFM) and then applied an affine transform to obtain the arousal and valence scores.
We fine-tuned the layer normalization parameters and affine matrices to minimize the squared error.

\noindent\textbf{Encoder-based models}: We report the scores of encoder-based methods~\cite{kang2025towards,louro2024merge}.
\cite{kang2025towards} integrated a music encoder with a chord recognition model and trained the model with both emotion classification and regression datasets~\cite{bogdanov2019mtg,soleymani20131000,zhang2018pmemo,aljanaki2017developing}. \cite{louro2024merge} trains a convolution-based music encoder from scratch on MERGE.

\noindent\textbf{Open models}: We included Qwen2-Audio-Instruct\footnote{\scriptsize\url{https://huggingface.co/Qwen/Qwen2-Audio-7B-Instruct}} and Phi-4-Multimodal\footnote{\scriptsize\url{https://huggingface.co/microsoft/Phi-4-multimodal-instruct}} as open models with parameter counts close to ours, and evaluated them in a zero-shot setting.

\subsection{Evaluation Protocol}
For evaluation, we followed the prompting procedure used in \cite{ma2025cmi}.
Predicted scores were extracted from generated responses using the method described in Sec.~\ref{sec:methods/feedback-alignment}.

The evaluation metric is the coefficient of determination $R^{2}$.
An $R^{2}$ of $0$ indicates performance equivalent to a method that always predicts the dataset mean, while higher values indicate more accurate emotion regression.

In our experiments, the score extraction succeeded for all samples and models, yielding a single arousal/valence score for each response.
Therefore, the reported $R^{2}$ scores are not affected by score-parsing failures and primarily reflect differences in numerical prediction accuracy.

\subsection{Results without MusicQA Fine-tuning}
\label{sec:experiments/results-without-musicqa}
We first evaluated emotion regression in an isolated setting without fine-tuning on MusicQA.
Table~\ref{tab:experiments/results-without-musicqa} summarizes the results.
We observed that, although the encoder-based models are specifically designed for music emotion regression, MusicFM probing is a strong baseline and outperforms the encoder-based methods in several cases.
This indicates that MusicFM provides effective feature representations for emotion regression.

With instruction tuning alone, our method does not achieve performance competitive with encoder-based conventional methods, and it also falls short of the probing baseline.
However, when feedback-driven alignment is applied after instruction tuning, the performance showed clear gains for both arousal and valence.
This suggests that instruction tuning alone is insufficient to fully leverage the capability of the music encoder, whereas feedback-driven alignment provides a more effective training signal by directly optimizing a reward based on the squared error.
Although probing or encoder-based models yield higher regression scores in some cases, they are limited to predicting emotion scores.
In contrast, our MusicLLM is not limited to score prediction and can also handle open-ended music QA, as we will show in the next section.

\subsection{Results with MusicQA Fine-tuning}
We then examine whether the model can retain general QA ability while improving regression performance.
Our goal here is not to replace task-specific regression-only probes, but to align a single MusicLLM for both emotion regression and general MusicQA.
As expected, models trained solely on MusicQA showed negative $R^{2}$ scores since MusicQA does not provide numerical supervision for emotion regression.
This is consistent with \cite{ma2025cmi}, where general QA supervision alone is insufficient to align MusicLLMs with precise emotion levels.

Instruction tuning improved arousal prediction, but yielded only limited gains for valence since valence annotations are more subjective than arousal~\cite{yang2008regression}.
However, applying feedback-driven alignment after instruction tuning enhanced both arousal and valence, yielding additional improvements for arousal and substantial gains for valence.
This suggests that feedback-driven alignment is important for MusicLLMs to predict emotion scores, even when instruction tuning is performed on a mixture of MusicQA and emotion regression data.
The lower regression scores may reflect the reduced proportion of emotion-regression supervision in the mixed objective.

To examine whether the model retains general QA ability, we also computed BLEU@4 (B-U)~\cite{papineni2002bleu}, METEOR (M-R)~\cite{banerjee2005meteor}, and ROUGE$_{\mathrm{L}}$ (R-L)~\cite{lin2004rouge} on MusicQA.
Results show that QA performance is preserved when emotion regression is included in instruction tuning, and remains comparable after feedback-driven alignment, even as regression performance improves, in our MusicQA evaluation.

Finally, we evaluated open models Qwen2-Audio and Phi-4-Multimodal in a zero-shot setting without additional fine-tuning.
Since their training data are not disclosed, it is unclear whether DEAM and MERGE were seen during training.
Both models report the use of preference optimization based on pairwise preference~\cite{rafailov2023direct} to improve instruction following and safety.
However, their $R^{2}$ scores remained negative in our experiments.
This is consistent with \cite{ma2025cmi}, indicating that task-specific adaptation, such as our instruction tuning and feedback-driven alignment, is important for accurate emotion regression.

\section{Conclusion}
In this work, we investigated whether MusicLLMs can learn emotion levels from music.  
We systematically compared instruction tuning and feedback-driven alignment.
In our experiments, instruction tuning yielded only limited improvements in emotion regression performance.
However, feedback-driven alignment with verifiable numerical feedback, particularly GRPO, substantially improved emotion regression accuracy, especially for valence, while preserving MusicQA capability in our evaluation.
These results suggest that numerical emotion supervision helps MusicLLMs capture continuous emotional dimensions, although our study is limited to specific datasets and model configurations.

To our knowledge, feedback-driven alignment using verifiable numerical feedback has not previously been explored for MusicLLMs.  
Since our method directly minimizes the score prediction error, it holds promise for extending MusicLLMs beyond emotion regression to other MIR tasks with verifiable numerical targets such as key and tempo estimation.
We hope our findings encourage further research on aligning MusicLLMs with broader aspects of music understanding.

\section{Generative AI Use Disclosure}
The authors used generative AI tools only to assist with grammar correction, improving the clarity of author-written text, and minor code debugging.
These tools were not used to generate scientific ideas, develop the methodology, design experiments, or produce experimental results.
All AI-assisted text and code were reviewed and verified by the authors, who take full responsibility for the submission.

\end{document}